\documentclass[12pt]{article} % single column, double spaced
\usepackage{ol2}
\usepackage[draft]{hyperref}
\usepackage{amsmath}

\begin{document}

%\twocolumn[ %% activate for two-column option

\title{
Amplification of surface plasmon polaritons in the presence of nonlinearity
and spectral signatures of threshold crossover}
\author{A. Marini$^1$, A.V.  Gorbach$^1$, D.V. Skryabin$^{1,*}$,  A. V. Zayats$^2$}
\affiliation{$^1$Centre for Photonics and Photonic Materials,
Department of Physics, University of Bath, Bath BA2 7AY, UK\\
$^2$Centre for Nanostructured Media, Queen's University of
Belfast, Belfast BT7 1NN, UK\\$^*$Corresponding author:
d.v.skryabin@bath.ac.uk}
\date{\today}
\begin{abstract}
We describe effects of nonlinearity on propagation of surface
plasmon polaritons (SPPs)  at an interface between a metal and an
amplifying medium of the externally pumped two-level atoms. Using
Maxwell equations we derive the nonlinear dispersion law  and
demonstrate that, the nonlinear saturation of the linear gain
leads to formation of stationary SPP modes with the intensities
independent from the propagation distance. Transition to the regime
of stationary propagation is similar to the threshold crossover in
lasers and  leads to narrowing of the SPP spectrum.
\end{abstract}

% ] %% activate for two-column option

Delivery of electromagnetic energy from macro- to nanoscales and
its direct generation in nanostructures are the challenging
problems of nanophotonics. They are particularly important
for plasmonics and metamaterials, which  promise
subwavelength localization and advanced control of
light in photonic nanocircuits. However,
metal structures typically suffer  from large intrinsic (Ohmic)
losses  hampering attractive applications. On the other hand, the
electromagnetic field confinement and enhancement associated with
surface plasmons have been proposed to control excitation of
active emitters and achieve a plasmonic laser (spaser)
\cite{Bergman2003,Zhel_Nat}. Propagating surface plasmon
polaritons (SPPs) at an interface with an amplifying dielectric
material have  been studied theoretically, see, e.g,
\cite{Nez_Opt_Expr,DeLion2008} and references therein, and
practically demonstrated using optically pumped dyes
\cite{Seid_PRL,Winter2006,Nog_PRL} and erbium doped glass
\cite{genov}. In particular, a distinct
threshold in the dependence of the output SPP intensity on gain
and the simultaneous significant narrowing of SPP spectra have
been reported inferring signatures of stimulated SPP emission.
\cite{Nog_PRL}.

Theory of SPPs interacting with an amplifying medium has been  so
far limited to the linear case
\cite{Bergman2003,Nez_Opt_Expr,Winter2006,Nog_PRL,DeLion2008}.
This approximation describes the SPP modes, where the linear
growth of the amplified SPPs is not balanced by the nonlinear
losses. In the linear theory  plasmons  with the real propagation
constant (in the SPP case) or the real frequency of the
oscillations (in the spaser case) can be found only  at the
threshold, where the linear loss is exactly balanced by the linear
gain \cite{Nez_Opt_Expr}. However, in practise, the gain-loss balance should
be maintained even above the threshold providing that nonlinearity and hence the
SPP intensity  are accounted for. This is because the  amplification of SPPs
is expected to be fully compensated by the nonlinear losses ensuring  existence of
SPPs with the real propagation constants.

Our main aim here is to present a theory of SPP
amplification  accounting  for saturation of the
linear gain by the nonlinear losses. This theory predicts
appearance of SPP modes with the stationary values of the
amplitudes reached above the threshold and reveals crucial role of
nonlinearity in shaping spectra of the excited SPPs.
We start our analysis from the time independent Maxwell equations for TM waves
\begin{eqnarray}
\label{tm1v2}
&& \partial_{zz} E'_x - \partial_{zx} E'_z=-D_x,\\
\label{tm2v2}
&& \partial_{zx} E'_x-\partial_{xx} E'_z=D_z,
\end{eqnarray}
where   $c$ is the vacuum
speed of light. $z$ is the coordinate along the interface and the
$x$ direction is orthogonal to it, both are measured in the units of
$1/k=\lambda_{vac}/(2\pi)$, where $\lambda_{vac}$ is the vacuum
wavelength. For the constitutive relation we assume
$\vec{D}=(\epsilon+\gamma|\vec{E'}|^2)\vec{E'}$, where
$\epsilon=\epsilon^{\prime}+i\epsilon^{\prime\prime}$ is the linear
permittivity and $\gamma=\gamma^{\prime}+i\gamma^{\prime\prime}$ is
the  nonlinear susceptibility. Below we use $\epsilon$ and $\gamma$
with subscripts $d$ and $m$  when referring to the dielectric
($x>0$) and the metal ($x<0$), respectively. The metal is assumed
linear, $\gamma_m=0$. Amplification in the dielectric is described
using the two-level model with relatively small SPP intensities.
Susceptibility of the  two-level atoms is $\chi(\delta)=\alpha
(\delta-i)/( \Gamma^2 +\delta^2)$ \cite{Boyd}, where $\delta$ is the
detuning of the SPP frequency from the atomic resonance frequency
 and $\Gamma=(1+|\vec
E'|^2/|E_*|^2)^{1/2}$ is the dimensionless intensity dependent
linewidth. $\delta$ and $\Gamma$ are both
normalized to the physical transition linewidth.
 $\alpha$ is the dimensionless  gain   per unit length.
We assume below that the SPP intensity is less than  the
saturation intensity $|E_*|^2$,  a particular value of the latter
depends on a material choice and is not important here.
The effect of the nonlinear saturation of the linear gain discussed below should
not be confused with and attributed to $|E_*|^2$.

We seek solutions of  Eqs. (\ref{tm1v2}-\ref{tm2v2}) in the form
$\vec E'(x,z)=E_*\vec{E}(x)\exp(i\beta z)$, where $\beta$ is the
propagation constant. Expanding $\Gamma$ and $\chi$ into the
Taylor series in $|\vec{E}|^2$ we find $\epsilon_d=\epsilon_b+\alpha(\delta+i)/ (1+\delta^2)$,
$\gamma_d=\alpha(i-\delta)/(1+\delta^2)^2$, where $\epsilon_b$ is
the dielectric constant of the background material hosting the
two-level atoms. For $\alpha>0$ we have linear gain and nonlinear absorption,
which  both are   maximal at the line center $\delta=0$.
We  are not taking into account the metal
dispersion, assuming that  it is negligible  relative to the
dispersion introduced by the strong two-level resonance.
Linear limit of the theory developed  below admits arbitrary complex $\beta$'s.
However nonlinear results require the stationarity of the SPP intensities with respect to the
propagation coordinate $z$, which is
achieved due to balance between all the loss and gain mechanisms.
Formally the balance condition is expressed as $Im\beta=0$.

The exponentially decaying for $x<0$ solutions can be readily found,
since the problem is linear inside the metal:
$E_x=B~e^{q_m x}$, $E_z=(iq_m/ \beta) B~e^{q_m x}$, $Re(q_m)>0$.
Here $q_m^2=\beta^2-\epsilon_m$ and $|B|^2$ is the SPP intensity
on the metal side of the interface. The system of equations we
need to solve for $x>0$ is
\begin{eqnarray}
&& \beta^2 E_x + i\beta \partial_{x} E_z= \left[\epsilon_d+\gamma_d\left(|E_x|^2+|E_z|^2\right)\right]E_x,
~~~~\label{tm1_stac}\\
&& i\beta \partial_{x} E_x-\partial_{xx} E_z=\left[\epsilon_d+\gamma_d\left(|E_x|^2+|E_z|^2\right)\right]E_z,
~~~~\label{tm2_stac}
\end{eqnarray}
The boundary conditions require continuity of $E_z$ and $D_x$ at $x=0$.
Assuming that $E_{x0}$, $E_{z0}$ are the  field components on the dielectric side of the interface
and knowing solutions inside the metal we  express the boundary
conditions using only the fields in the dielectric
\begin{eqnarray}
 \beta\epsilon_m E_{z0}=iq_m\left[
\epsilon_d+\gamma_d\left(|E_{x0}|^2+|E_{z0}|^2\right)
\right]E_{x0},
\label{BC1}
\end{eqnarray}
Solving Eqs. (\ref{tm1_stac},\ref{tm2_stac})
perturbatively under the assumption that nonlinear terms $\gamma_d|\vec E|^2$ are small
we found
\begin{eqnarray}
\label{exlin}
E_x=Ae^{-q_d x}\{1+w_x\gamma_d|A|^2e^{-2xReq_d}+O(|\gamma_d|^2)\},\\
E_z=\frac{q_d}{i\beta}Ae^{-q_d x}\{1+w_y\gamma_d|A|^2e^{-2xReq_d}+O(|\gamma_d|^2)\}.
\label{ezlin}
\end{eqnarray}
Here $q_{d}^2=\beta^2-\epsilon_{d}$, $Re(q_d)>0$, and $w_{x,y}$
are  some constants not shown here.
$|A|^2$ characterizes the SPP intensity on the dielectric side of
the interface. Substituting the above solutions into  Eq.~(\ref{BC1}), we find the dispersion law for SPPs
accounting for nonlinearity, losses in metal and gain in the
dielectric
\begin{eqnarray}
\label{beta_A_pert}
\epsilon_mq_d+\epsilon_dq_m=\gamma_d|A|^2F+O(|\gamma_d|^2).\end{eqnarray}
The constant $F$ is  given by
\begin{eqnarray}
\label{feq} && F\equiv\left(\frac{|q_d|^2}{\beta^2}+1\right)\times\\
\nonumber && \frac{q_d\epsilon_m+q_m\epsilon_d+2Re(q_d)(\beta^2\epsilon_m/\epsilon_d-q_mq_d)}{4Re(q_d)(Re(q_d)+q_d)}.
\end{eqnarray}
Dispersion of nonlinear  SPPs in the absence of
gain and loss has been previously derived, e.g., in Ref. \cite{Miha_OPL}.

For $|A|=0$, Eq. (\ref{beta_A_pert}) transforms into a well known
linear dispersion law for SPPs, which is readily resolved with
respect to $\beta$:
$\beta=\beta_l\equiv\sqrt{\frac{\epsilon_{m}\epsilon_d}{\epsilon_{d}+\epsilon_m}}$.
Practically, SPPs can be excited either directly by the two-level
emitters (dye molecules, quantum dots, etc) or externally coupled
into the system, e.g., through a prism or a grating. Depending on
the spatial variations of the emitter density and on the
excitation type  (optical, electric or chemical pumping), the
population inversion and hence the gain coefficient $\alpha$ may
vary with the distance from the interface, see, e.g.
\cite{DeLion2008}. We will not consider these effects in order to
focus on the role of nonlinearity and to derive closed analytical
expression for the nonlinear SPP dispersion.
To obtain physical estimates for our dimensionless calculations,
we used $\epsilon_b=1.8$ (polymer) and $\epsilon_m=-15+i0.4$
(silver at $\lambda_{vac}=530$ nm). For these parameters, without
the resonant atoms ($\alpha=0$) the characteristic SPP propagation
distance is $(kIm\beta_l)^{-1}\simeq 30\mu$m.

If we neglect the effect of nonlinearity, the condition
$Im\beta_l=0$ corresponds to the lossless SPP propagation and it
is achieved at the gain threshold $\alpha=\alpha_0$:
$\alpha_0=\frac{1}{2\epsilon''_m}[\left(|\epsilon_m|^2-2\epsilon''_m\epsilon_b\delta\right)-
 \{\left(|\epsilon_m|^2-2\epsilon''_m\epsilon_b\delta\right)^2
-4(\epsilon''_m)^2(\epsilon_b)^2\left(1+\delta^2\right)\}^{1/2}]$.
As expected, the lowest gain $\alpha_0=\alpha_{min}$ required for
the lossless propagation happens  at the exact resonance
($\delta=0$): $\alpha_{min}\simeq 0.00575$. For example for
$\alpha=1.5\alpha_{min}$ and $2\alpha_{min}$ the characteristic
SPP  gain length $(kIm\beta_l)^{-1}\simeq
65\mu$m and $30\mu$m, respectively.
The gain length is defined as the
propagation distance at which the SPP amplitude increases by a factor $e$.
Intensity of the nonstationary
($Im\beta\ne 0$) SPPs in the linear case  can be easily calculated: $I_l\sim
e^{-2zIm\beta_l}$. The typical dependence of $I_l$ on $\delta$ is
shown in Fig. 1(a) (line 2). Lines 2,3 in Fig. 1(b) show how the
full width half maximum (FWHM) of $I_l(\delta)$ varies with the gain
parameter $\alpha$ for two different propagation distances. One
can see that the spectrum quickly narrows as the gain is increased, but  kept
below the threshold ($\alpha<\alpha_{min}$). Close to and above the
threshold  the narrowing continues, but at a much slower pace.
With the increase of the propagation distance and for the fixed
gain, the spectrum also narrows (cf. lines 2 and 3 in Fig. 1(b)),
since the spectral components of SPP modes near the line center
are stronger amplified and hence become dominant.

The influence of nonlinear effects on the SPP propagation constant
can be derived by solving Eq. (\ref{feq}) with respect to $\beta$ and demanding $Im\beta=0$.
We assume, that the right-hand side of Eq. (\ref{feq}) can be treated as a perturbation and find:
\begin{equation}
\label{betanon}
\beta=\beta_l+\beta_{nl}|A|^2,~\beta_{nl}\equiv
\frac{\beta_l\gamma_d}{2\epsilon_d^2}\frac{q_d(|q_d|^2+\beta_l^2)}{Re(q_d)+q_d},
\end{equation}
where $q_{d,m}$ inside $\beta_{nl}$ are calculated for
$\beta=\beta_l$.
Above the threshold ($\alpha>\alpha_{min}$), the linear growth of the
SPP intensity is saturated by the nonlinear absorption and as a result the intensity
quickly  attains a stationary value.
To calculate the spectral and other characteristics of the stationary
(saturated) SPPs, we use Eq. (\ref{betanon}) and impose the condition $Im\beta=0$.

In the
linear approximation, $Im \beta_l=0$ implies $\alpha=\alpha_0$ (see above) and
the corresponding real propagation constant is
$\beta_l(\alpha_0)\equiv\beta_{0}$. Expanding $\beta$ in Eq. (\ref{betanon}) into the
Taylor series in $(\alpha-\alpha_0)$, i.e. close to the threshold for a given $\delta$, we
find
$\beta=\beta_0+(\alpha-\alpha_0)\partial_{\alpha}\beta_l+\beta_{nl}|A|^2$.
Then $Im\beta=0$ gives the  intensity $I_s\equiv |A_s|^2$ of
the stationary SPPs:
\begin{equation}\label{int}
I_s={(\alpha-\alpha_0)\mathrm{Im}\partial_{\alpha}\beta_l/( -\mathrm{Im}\beta_{nl})},
\end{equation}
where $\partial_{\alpha}\beta_l$ and $\beta_{nl}$ are  calculated
for  $\alpha=\alpha_0$. Fig. 2 shows dependencies of $I_s$ on gain
for several values of $\delta$. Naturally, one can see that for
higher gain,  SPPs within the wider
frequency interval around the resonance cross the threshold,
leading to the gradual broadening of the spectra of the stationary SPPs (see Line 1 in Fig. 1(b)).

Fig. 1(a) compares the spectral intensity profiles
of the linear SPPs below threshold and of the nonlinear saturated
ones above the thresholds. Also, Fig. 1(b) compares
the linewidth of these two SPP families. One can see that the spectra of
stationary SPPs above the threshold are much narrower for small
deviations from $\alpha_{min}$, than spectra of the linear SPPs. For propagation
distances of few gain lengths $(kIm\beta_l)^{-1}$ (e.g.,
$100\mu$m) we shall expect that the SPPs
should achieve their stationary saturated intensities. Therefore, the threshold
crossover, if observed at sufficiently long distances, should be
accompanied by a marked spectral narrowing, which has indeed been
reported in the recent experiments \cite{Nog_PRL}. Note the
obvious qualitative agreement between  our Figs. 1(a) and 2 and
the experimental Figs. 2(a) and 2(b) in Ref. \cite{Nog_PRL}.

For the propagation constant of the stationary saturated SPPs  we thus have
$\beta_s=\beta_0+(\alpha-\alpha_0)\mathrm{Re}\partial_{\alpha}\beta_l+I_s\mathrm{Re}\beta_{nl}$.
Fig. 3 compares real and imaginary parts of the above $\beta_s$ with $\beta_l$.
It shows that  there exists a marked difference of the dependencies
of the propagation constant from the gain parameter in the
linear and nonlinear cases.
We  have independently cross-checked the
SPP profiles and propagation constants using numerical shooting
method applied directly to the Maxwell equations. Good
agreement between analytical and numerical
results for $\alpha<2\alpha_{min}$ made it unnecessary to present
the numerical results in the context of this paper.

In summary, we have presented a theory of SPP amplification in the
presence of nonlinear gain saturation as it  happens in the
real-world systems. We have calculated saturated intensities of
SPPs above the amplification threshold and quantitatively
described transition from the broad Lorentzian spectra of the
nonstationary amplified SPPs to the narrow spectra of the stationary
saturated SPPs above the threshold. Our analytical results are
expressed in terms of complex dielectric permittivities of the
materials involved and can be readily used in design of plasmonic
devices.

The work of A. Zayats has been  supported in part by EPSRC (UK) and by EC
FP6 project PLASMOCOM. The authors are grateful to A. Krasavin for
discussions.

%\bibliographystyle{ol}
%\bibliography{Plasmon_Gain}

\begin{thebibliography}{10}
\newcommand{\enquote}[1]{``#1''}

\bibitem{Bergman2003}
D.~J. Bergman and M.~I. Stockman, "Surface Plasmon Amplification by Stimulated Emission of Radiation:
Quantum Generation of Coherent Surface Plasmons in Nanosystems," Phys. Rev. Lett. \textbf{90}, 027402
  (2003).



\bibitem{Zhel_Nat}
N.~I. Zheludev, V.~Prosvirnin, N.~Papasimakis, and V.~A. Fedotov, "Lasing spaser", Nature
  Photonics \textbf{2}, 351  (2008).

\bibitem{Nez_Opt_Expr}
M.~P. Nezhad, K.~Tetz, and Y.~Fainman, "Gain assisted propagation of surface plasmon polaritons on planar metallic waveguides",
Opt. Express \textbf{12}, 4072 (2004).

\bibitem{DeLion2008}
I.~De~Lion and P.~Berini, "Theory of surface plasmon-polariton amplification in planar structures incorporating dipolar gain media", Phys. Rev. B \textbf{78}, 161401(R) (2008).

\bibitem{Seid_PRL}
J.~Seidel, S.~Grafstrom, and L.~Eng, "Stimulated Emission of
Surface Plasmons at the Interface between a Silver Film and an Optically Pumped Dye Solution", Phys. Rev. Lett. \textbf{94},
  177401 (2005).

\bibitem{Winter2006}
G.~Winter, S.~Wedge, and W.~Barnes, "Can lasing at visible wavelengths be achieved
using the low-loss long-range surface
plasmon-polariton mode?", New J. Phys. \textbf{8}, 125
  (2006).

\bibitem{Nog_PRL}
M.~A. Noginov, G.~Zhu, M.~Mayy, B.~A. Ritzo, N.~Noginova, and V.~A. Podolskiy,
"Stimulated Emission of Surface Plasmon Polaritons",   Phys. Rev. Lett. \textbf{101}, 226806 (2008).

\bibitem{genov}
M. Ambati, S.H. Nam, E. Ulin-Avila, D.A. Genov, G. Bartal, and X. Zhang,
Nano Lett. {\bf 8}, "Observation of Stimulated Emission of Surface Plasmon Polaritons", 3998 (2008).

\bibitem{Boyd}
R.~W. Boyd, \emph{Nonlinear Optics} (Academic Press, 2003).

\bibitem{Miha_OPL}
D.~Mihalache, G.~I. Stegeman, C.~T. Seaton, E.~M. Wright, R.~Zanoni, A.~D.
  Boardman, and T.~Twardowski, "Exact dispersion relations for transverse magnetic polarized guided waves at a nonlinear interface",
  Opt. Lett. \textbf{12}, 187 (1987).


\end{thebibliography}

\newpage

%\end{document}
\newpage
%\centerline{Figure captions}

\begin{figure}
%\centerline{
%\includegraphics[width=0.22\textwidth]{fig1a.eps}
%\includegraphics[width=0.22\textwidth]{fig1b.eps}
%}
\caption{(Color online)(a) Normalized SPP intensity as a function of detuning
$\delta$. Line 1 is the
spectrum of stationary (saturated) SPP, $I_s(\delta)$: $\alpha=1.1\alpha_{min}$.
Line 2 is the spectrum of the amplified linear SPP, $I_l(\delta)$,  below the
threshold (propagation distance  $60\mu$m,
$\alpha=0.8\alpha_{min}$).   (b)
FWHM of the SPP spectra vs gain. Line 1 is for the
stationary saturated SPPs, $I_s$.
Lines 2 (propagation distance  $60\mu$m) and 3 (distance $100\mu$m) are for the nonstationary linear
SPPs, $I_l$.} \label{spectr}
\end{figure}

\begin{figure}
%\centerline{
%\includegraphics[width=0.35\textwidth]{Intensity_figure.eps}}
\caption{(Color online) The dependence of the intensity of the stationary SPPs
above the threshold on the gain parameter $\alpha/\alpha_{min}$.
Line 1 corresponds to $\delta=0$, Line 2 to $\delta=0.4$ and Line
3 to $\delta=0.8$.} \label{intensity}
\end{figure}

\begin{figure}
%\centerline{
%\includegraphics[width=0.22\textwidth]{fig3a.eps}
%\includegraphics[width=0.22\textwidth]{fig3b.eps}
%}
\caption{(Color online)
The dependence of the real (a) and imaginary (b) parts of the SPP propagation constants on the
gain parameter for $\delta=0$. Solid red lines
correspond to the stationary nonlinear SPPs ($\beta_s$)
and the dashed  black lines to the linear SPPs ($\beta_l$).}
\label{bet}
\end{figure}

\newpage

\setcounter{figure}{0}

\begin{figure}
\centerline{
\includegraphics[width=0.45\textwidth]{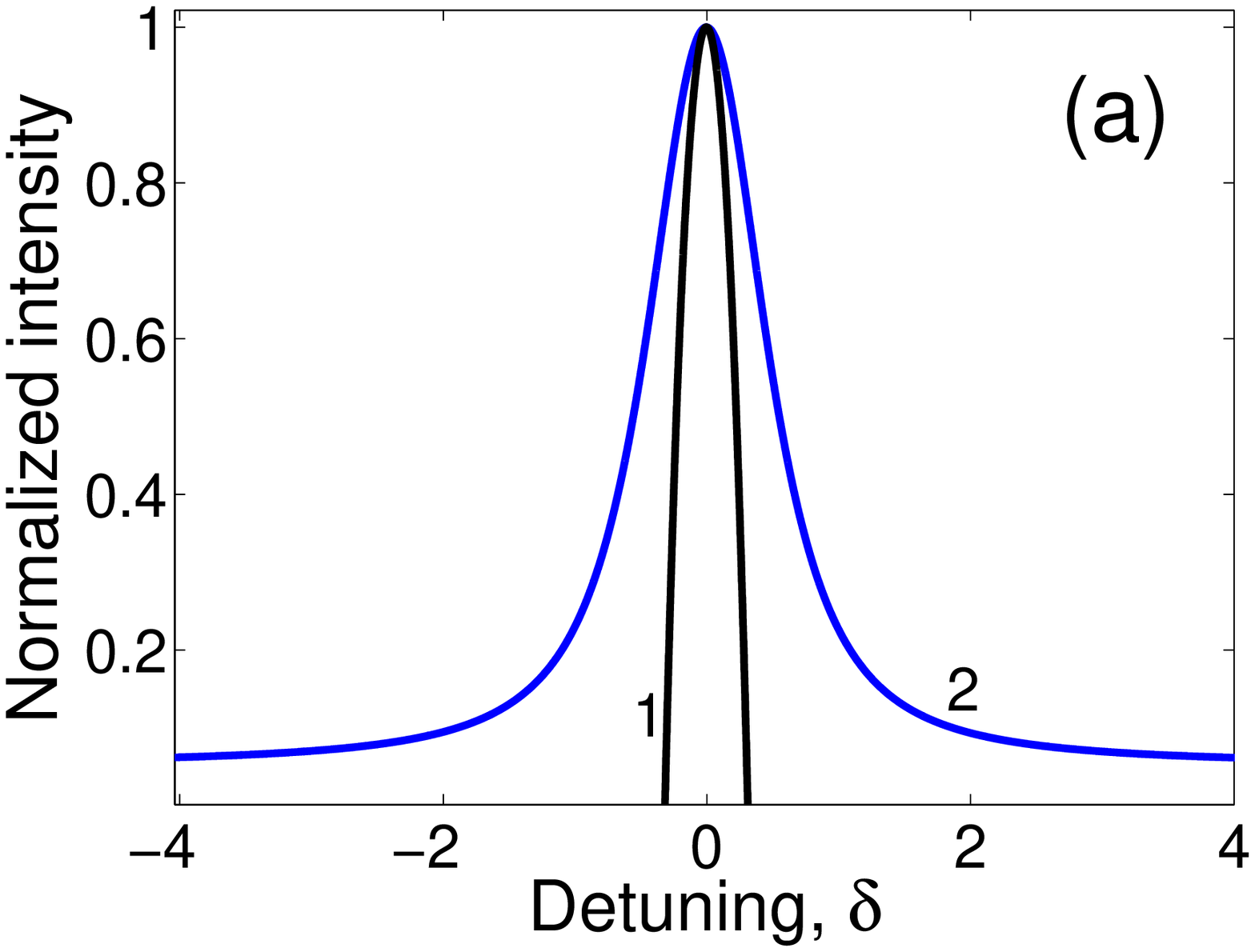}
\includegraphics[width=0.45\textwidth]{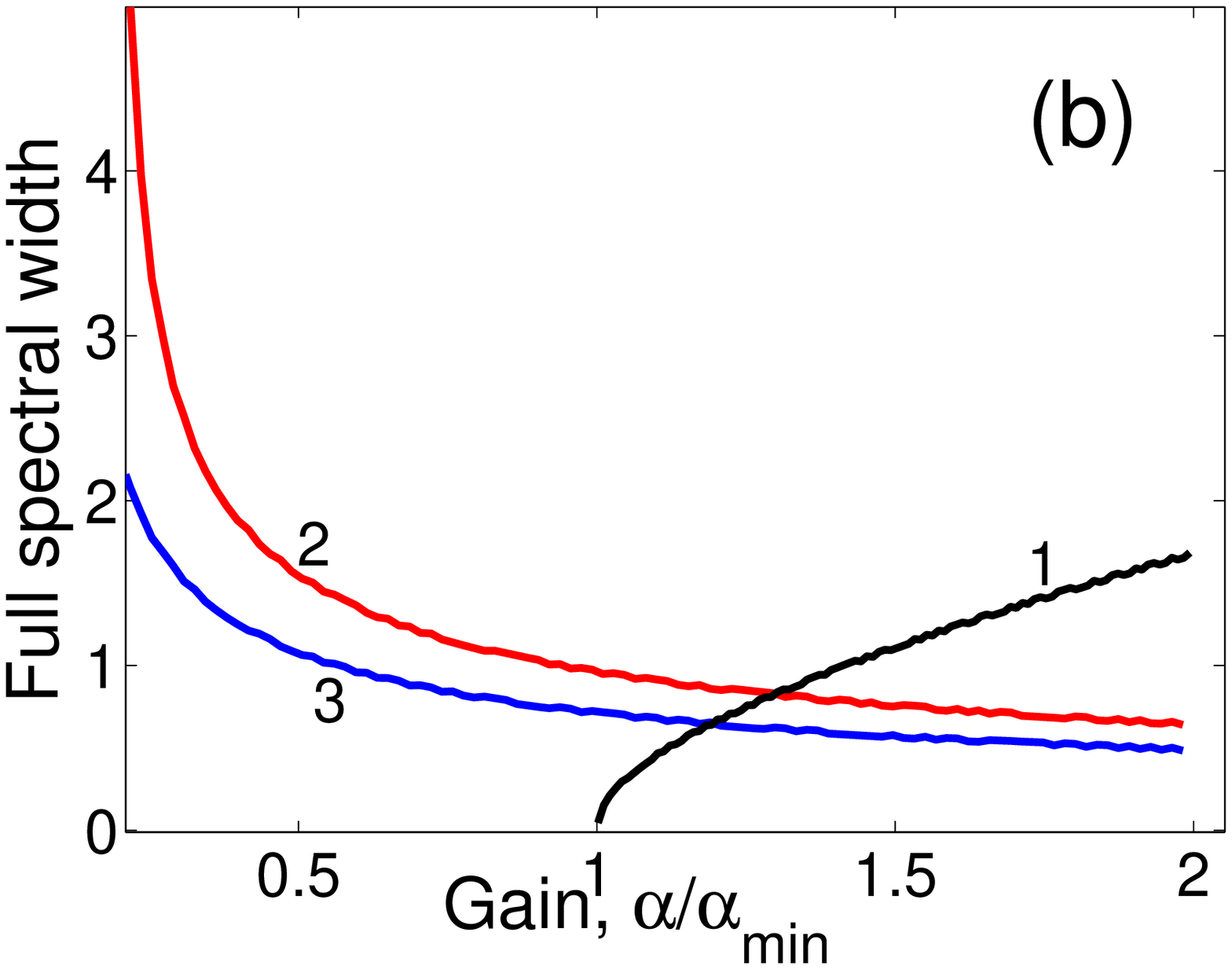}
} \caption{} \label{spectr}
\end{figure}

\begin{figure}
\centerline{
\includegraphics[width=0.8\textwidth]{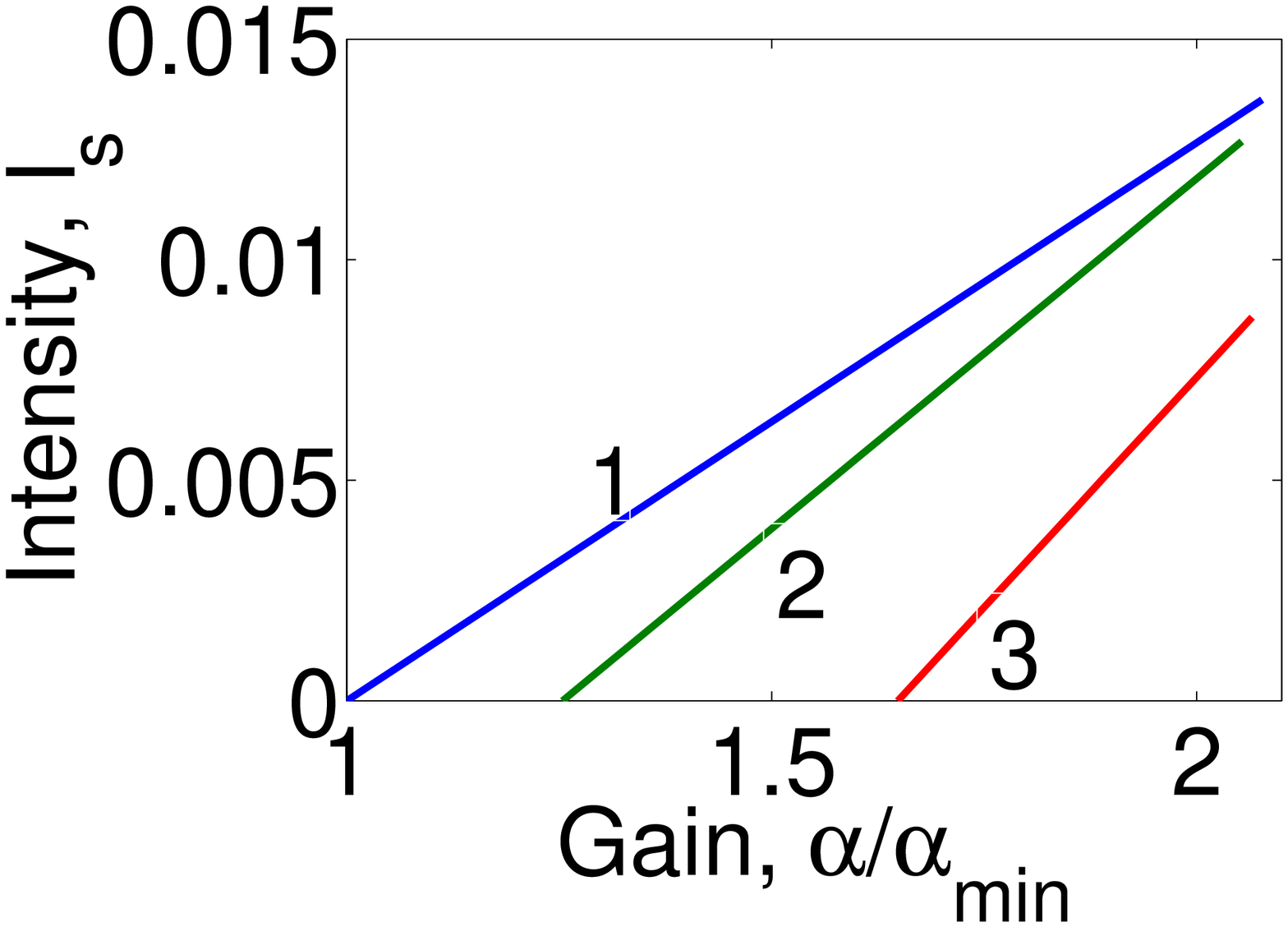}}
\caption{} \label{intensity}
\end{figure}

\begin{figure}
\centerline{
\includegraphics[width=0.45\textwidth]{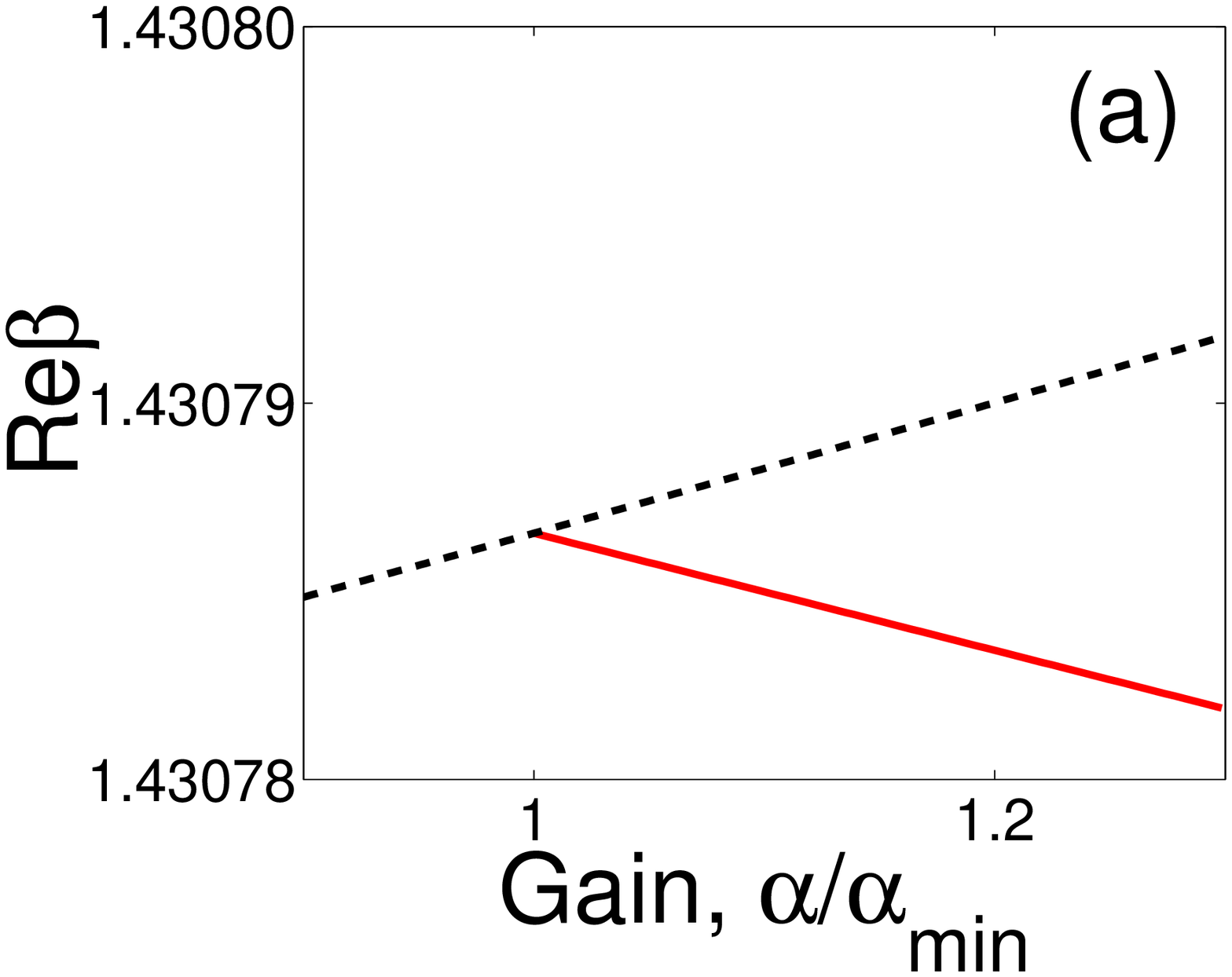}
\includegraphics[width=0.45\textwidth]{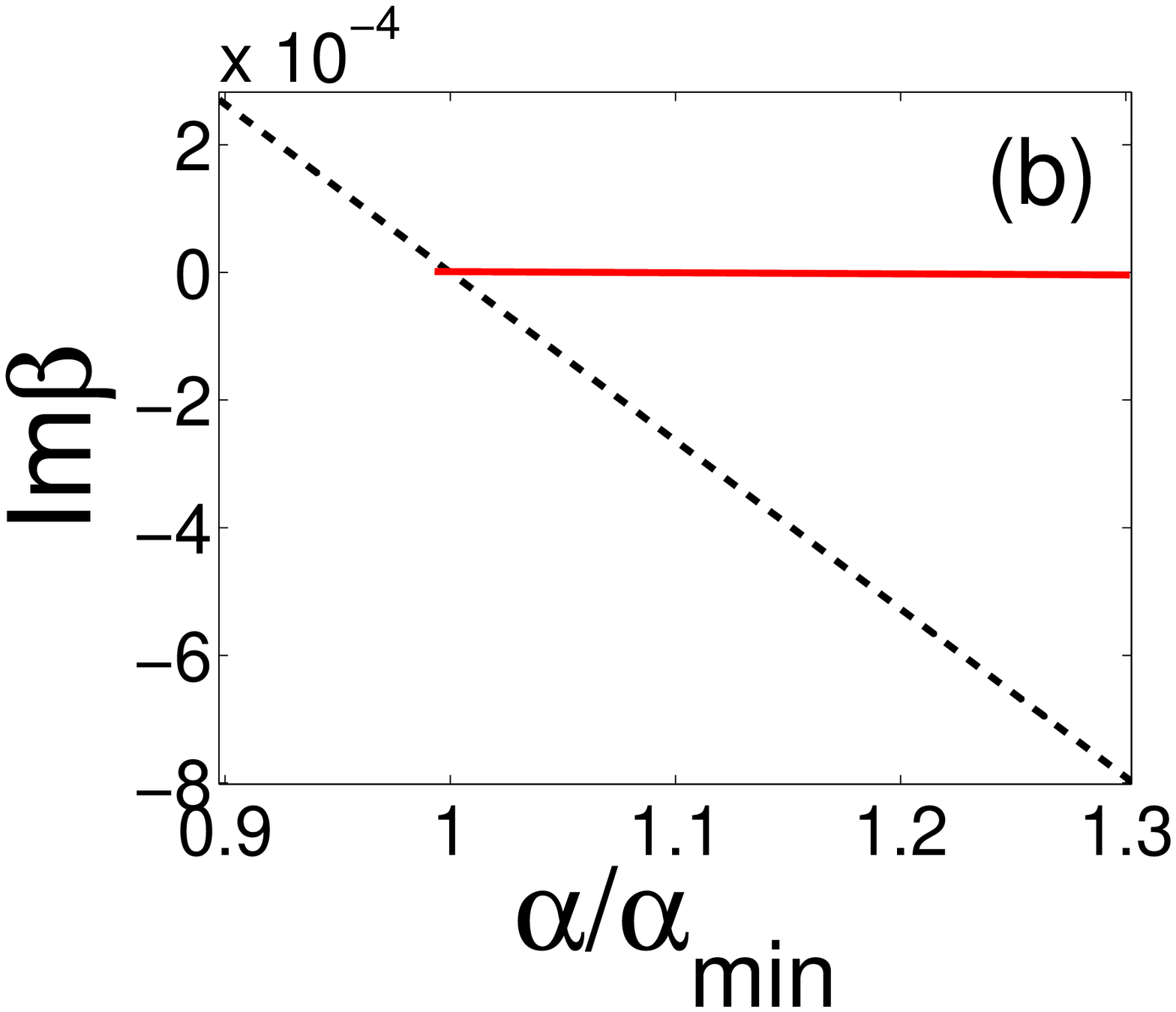}
}
\caption{}
\label{bet}
\end{figure}

\end{document}